# Suppressed Polaronic Conductivity induced Sensor Response Enhancement in Mo doped $V_2O_5$ Nanowires

Anakha Anson, Dipanjana Mondal, Varsha Biswas, Kusuma Urs MB, Vinayak Kamble*

School of Physics, Indian Institute of Science Education and Research Thiruvananthapuram, India 695551.

**Abstract**
In this paper, we show the direct correlation between suppression of polaronic oxygen vacancy defect ($V_o$) density and gas sensor response of 1 at% Mo doped $V_2O_5$ (MVONW) nanowires. Doping 1 at% $Mo^{5+}$ leads to substitution at the $V^{5+}$ site in $V_2O_5$ nanowires (VONW) and thereby reduction in $V_o$ defects. This in turn affects the charge carrier hopping sites and subsequently enhances the sensor response at lower temperatures (<320 °C). The $Mo^{5+}$ dopants lead to the lowering of Fermi energy ($E_F$) towards valence band maxima due to reduced $V_o$ donor density. The polaron suppression is confirmed with activation energy of polaron hopping, increasing from 195 meV to 385 meV in VONW and MVONW respectively. As a result, the response to ethanol gas enhanced as the depletion width is widened for the given cross-section of the nanowires. This may lead to large depletion controlled cross-sectional area and thereby better sensitivity. At about 350 °C VONW show change in slope of resistance vs temperature (MIT) which is not observed in case of MVONW. This is attributed to presence of enhanced non-stoichiometry of V ion resulting in metallic behaviour and accompanied with sudden rise in sensor response at this temperature. Moreover, the absence of MIT may be attributed to lack of such sudden rise in response in MVONW.

*Email: kbvinayak@iisertvm.ac.in

## I. Introduction

Semiconducting Metal Oxides (SMO), are the most commonly used sensing material and remain so[1-3]. It works on the variation of surface electrical conductance in the presence of a gaseous environment. Medium to wide bandgap oxide semiconductors are an excellent choice for gas sensing material due to the presence of intrinsic donors/acceptors in the bandgap and large changes in surface potentials[4-7]. Due to this, there will be either electrons in the conduction band or holes in the valence band which makes it possible to have surface gas reactivity even at room temperature [4,5]. SMO as gas sensors offer advantages like low cost, easy fabrication, ease of use, and detection of a wide variety of gases.

Vanadium pentoxide is one such sensor material with promising gas-sensing properties. $V_2O_5$ belongs to a class of Transition Metal Oxides (TMOs) with '$d$' metal valence orbitals. They exhibit multiple oxidation states owing to the small energy difference between the $d^n$ configuration and either $d^{n-1}/d^{n+1}$ configuration. As a result, it is much easier to engineer defects in the lattice that could increase conductivity[8,9]. TMOs bring special features to gas sensing due to their unique surface properties and high catalytic activity. Thus, specific catalytic activity may be exploited to design selective sensors for a particular analyte [8,10]. Vanadium, a transition metal with a [Ar] $3d^34s^2$ electronic configuration, forms numerous and frequently complicated compounds because of its variable oxidation states. It exhibits four oxidation states: +2, +3, +4, and +5.

Among various oxide forms of vanadium, vanadium pentoxide ($V_2O_5$) is thermodynamically most stable and is an n-type semiconductor with low mobility[11]. The n-type conductivity arises because of excess electrons contributed by oxygen vacancies in the lattice. $V_2O_5$ crystallizes in an orthorhombic system and has a layered structure where van der Waal's force act between the layers. In $V_2O_5$, oxygen vacancies are the most dominant defects. This point defect plays an important role in the conductivity of $V_2O_5$ by inducing the formation of compensating electron polarons that act as charge carrier[12]. The conductivity of $V_2O_5$ stated in literature is due to the "hopping" mechanism of polarons, i.e., an electron in $V_2O_5$ hopping from a $V^{4+}$ center to an adjacent $V^{5+}$ site[13].

Doping is one of the main methods employed to improve the performance of resistive-type SMO sensors[14]. Doping normally involves replacing a metal ion in the crystal lattice of the metal oxide with the dopant ion. The process of doping alters the electronic levels, electrical properties, structural alignments of grains, size, shape, surface morphologies of nanoparticles, and optical, chemical, or catalytic activities of the metal-oxide nanocomposites[12]. The dopants may occupy a substitutional site (of comparable ionic radius) or may occupy the interstitial site (smaller atomic radius like Li ion). The layered structure allows easy intercalation of such smaller ions with the layer[9]. Intercalation of metal ions or the gas phase species into the layered structure of vanadium oxide (metal cations doping) is an effective tool to enhance the sensitivity and improve the functionalities of $V_2O_5$ based sensors [10,15,16]. Depending on the ionization of the substituent metal ion, it may either act as a donor or acceptor. In either case, this metal ion doping helps in tuning the chemical potential of the semiconductor thus affecting its electrical conductivity. This could have a doubly induced effect on sensitivity i.e., preferred selective reaction towards certain chemical compounds and/or changes in the surface depletion of the oxide. Molybdenum is also a transition metal that shows variable oxidation states such as +4, +5, and +6. It has a high probability of replacing vanadium into a $V_2O_5$ lattice via substitutional doping because $Mo^{6+}$ ion (r = 59 pm) has a radius similar to that of $V^{5+}$ ion (r = 54 pm).

So far, various literature has reported Mo-doped $V_2O_5$ nanostructures to have enhanced various properties like electrochromism[17], charge storage [18,19], and sensing response [20], however, the enhancement has not been explained to sufficient details. Therefore, 1 at% Mo doping into $V_2O_5$ nanowires has been attempted in this paper. We observed that Mo still takes $Mo^{5+}$ valence and therefore reduces the non-stoichiometry of $V^{4+}$ to $V^{5+}$ by forcing a reduction in oxygen vacancy defect which neutralizes the charge reduction due to $V^{4+}$. The sensing response measurements of $V_2O_5$ nanowires were performed using ethanol ($C_2H_5OH$) as the target gas. Ethanol is a reducing gas that has significant industrial and pharmaceutical applications. As its long-term exposure leads to various ill effects on the health system, detecting ethanol is very important in both indoor and outdoor environments [16]. Besides, ethanol is one of the main volatile compounds released by many food items at the onset of degradation. Therefore, it is important to detect the low concentration of ethanol to prevent food from degrading.

## II. Experimental

The $V_2O_5$ nanowires were prepared by a hydrothermal method described elsewhere [21]. 1 at% Mo-doped $V_2O_5$ powder was synthesized using the reaction between Ammonium Metavanadate [$NH_4VO_3$] and Ammonium Molybdate [$(NH_4)_6Mo_7O_{24}$] that exploits the unique interaction between $V_2O_5$ and $MoO_3$ owing to the similarity of ionic radii and the structures in their highest oxidation state [22]. Hereafter the undoped and 1 at% Mo doped $V_2O_5$ Nanowires are referred to as VONW and MVONW.

The prepared VONW and MVONW nanowire samples were characterized using Bruker X-ray diffractometer having a copper $K_\alpha$ source of 1.5418 Å in the range 5º to 90º and in step size of 0.017º. The morphological properties were studied by using a Nova Scanning Electron Microscope (SEM) using 10 kV. The X-ray Photoelectron Spectra (XPS) and Ultraviolet Photoelectron Spectroscopy spectra (UPS) were obtained on an Omicron Nanotechnology-made XPS system with Mg Kα X-ray source (1253.6 eV) for XPS and He II source (21.2 eV). The I–V characteristics and gas sensing properties of the VONW and MVONW were measured by two probe methods using Keithley 6517B Electrometer. The details of the gas sensing system can be found elsewhere [23,24]. The dispersion of powder samples in ethanol was drop-casted over a gold IDE (Inter-Digitated Electrodes). The temperature was controlled using a temperature controller connected to a high-temperature stage (LINKAM, UK). A DC voltage sweep from -5 to 5 V was applied to the samples and the corresponding current values were recorded at different temperatures. The gas sensing tests were carried out by monitoring the electrical resistance changes of the nanowires towards ethanol at different operating temperatures.

## III. Results and Discussion
### A. Structural and morphological studies

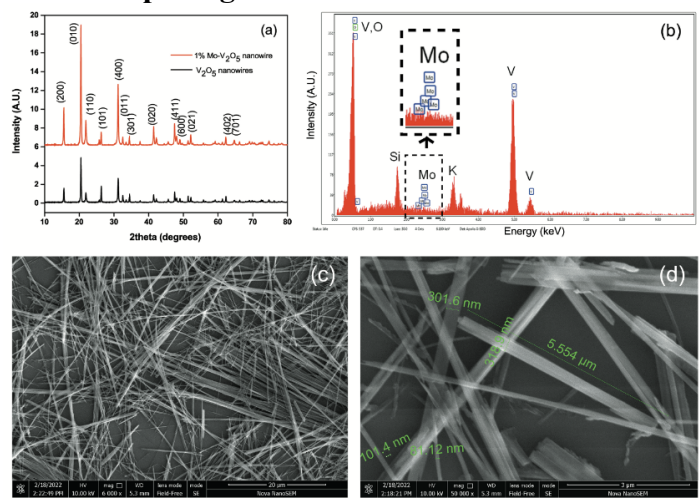

**Figure 1.** The (a) X-ray diffraction patterns of undoped and 1% Mo-doped $V_2O_5$ nanowires. (b) The EDS spectrum shows the Mo signal from nanowires magnified inset. The SEM images recorded at (c) low and (d) high magnification show a long aspect ratio of the nanowires with typical widths ~ 100 nm and lengths >10 μm.

As XRD patterns are shown in Figure 1(a), both Mo-doped and undoped samples are orthorhombic $V_2O_5$, and there is no signal for molybdenum oxide or any other impurity. This implies that Mo has entered into the $V_2O_5$ matrix via substitution or interstitial doping. Both the patterns demonstrate strong and sharp peaks, suggesting the high crystallinity of samples. It may be also seen that the Mo-doped sample exhibited slightly broader XRD peaks than the undoped sample, which suggests the crystal size of the former is comparatively smaller. It may be seen from Figure S1 in the supporting information section that some peaks have shifted towards either side whereas others directly align with undoped peaks. This confirms that the crystal structure has been distorted as some $V^{5+}$ ions have been replaced by Mo ions. The substitution of Mo ions into the bulk $V_2O_5$ without affecting the long-range periodic structure of orthorhombic $V_2O_5$ may thus be confirmed. The ionic radii of $V^{5+}$, $Mo^{5+}$ and $Mo^{6+}$ ions are 0.49, 0.6, and 0.55 nm in tetrahedral coordination.

The Energy Dispersive Spectroscopy (EDS) measurements on the doped nanowires did show the presence of Mo as shown by the spectrum presented in Figure 1(b). However, due to the low concentration of Mo which is below the resolution of EDS, the quantification is found to vary from point to point. Nevertheless, it may be used to confirm the presence of Mo throughout the nanowire, and Mo : V ratio was between 0-1%. The small boxes with letters marked V, O, Mo etc denotes vanadium, oxygen and molybdenum characteristic X-ray lines. These are identified depending on the threshold intensity by the quantification software. It may be seen from Figure 1(c-d) that the sample is in the form of nanowires with lengths ranging between 10 μm to 30 μm and widths ranging from 100 nm to 300 nm. The nanowires are sufficiently long and the small width gives it a large surface area to volume ratio which can increase the sensitivity of the sensor. When compared, it may be seen that doped nanowires have greater lengths ranging between 10 μm to 100 μm, and widths varied between 80 nm to 350 nm.

**B. Electron spectroscopy for chemical and electronic structure analysis.**

As shown in Figure 2(a), the core levels of V 2p and O 1s overlap in binding energy, making the analysis of XPS spectra challenging. However, the difference between the two core levels' spectra is clearly evident. It may be seen that the O 1s region shows a significant change from the undoped to Mo-doped sample. Also, the shape of the V 2p region, precisely the peak asymmetry shows a marked change. The peak asymmetry arises from the contribution of different chemical shifts i.e., $V^{5+}$ and $V^{4+}$ ions, and their relative contributions. Clearly, the undoped sample has a relatively higher intensity for the $V^{4+}$ region (colored in yellow at $3/2$ for 516 eV and $1/2$ at 523 eV) compared to $V^{5+}$ peaks (colored in pink at $3/2$ for 517 eV and $1/2$ at 524 eV) [25,26]. Thus, the Mo-doped sample shows a higher $V^{5+}$ contribution as can be seen from the peaks marked in Fig. 2(a) and (b).

The native O 1s is also asymmetric due to a contribution from O vacancies peak (green color) at about 530.5 eV in addition to the lattice oxygen $O^{2-}$ peak (purple/violet color) at 529.5 eV[25-27]. Thus, it may be concluded that the O vacancies have reduced in Mo doped sample. Thereby, commensurately $V^{4+}$ contribution is also decreased as these are charge compensatory defects for the oxygen vacancies. Similarly, the O 1s peak shows another peak at a high binding energy of 532 eV (turquoise color) which could be ascribed to surface-adsorbed oxygen species[28,29]. However, this adsorption-induced peak is much more significant in Mo-doped NWs which may indicate improved adsorption active sites and hence may be beneficial for sensing purposes.

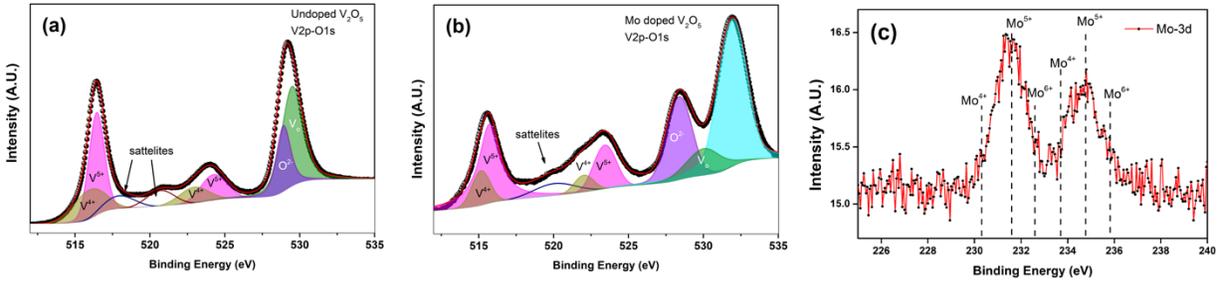

**Figure 2.** Comparison of the V 2p and O 1s overlapping core level XPS spectrum of (a) undoped and (b) Mo-doped NWs. (c) 3d core level spectrum of Molybdenum.

Mo 3d core level spectrum is rather noisy due to the very low concentration of Mo as shown in Figure 2(c). However, upon examination, it is observed at binding energies matching with the $Mo^{5+}$ region (233 and 235 eV)[30] which agrees with the assumption that $Mo^{5+}$ may be replacing $V^{5+}$ in the lattice. Thus, the vanadium non-stoichiometry in Mo-doped $V_2O_5$ nanowires is subsequently partially restored upon introducing Mo ions.

It may be seen from Figure 2(d) that the two samples showed different valence band regions (enlarged in Figure 3(d)) upon comparison. However, the SEC of the two samples occurs at nearly the same binding energy (16.4 eV as seen in Figure 3(c)). This could be because of the small change in work function that is adjusted in change in VBM as seen in Figure 3(d). It should be noted that in Eq (9) if SEC and hv are the same, the change in ϕ is due to change in VBM. Further, it is also possible that the small difference in the work function may be masked by a large thermal broadening of secondary electron binding energy cut-off. As the VBM of the Mo doped sample moves towards $E_f$ (VBM reduces by 0.2 eV as seen in Figure 3(d)., the WF increases for Mo doped $V_2O_5$ NW by about 0.2 eV.

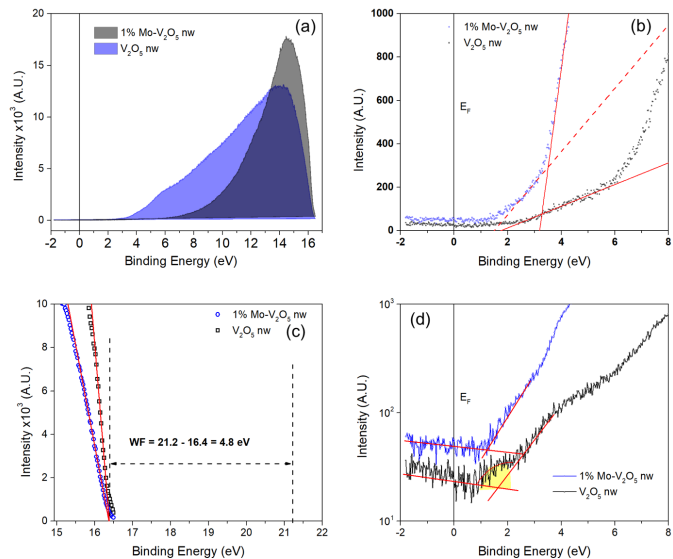

**Figure 3.** (a) The comparison of UV Photoelectron spectra (UPS) of the undoped and 1% Mo doped $V_2O_5$ nanowires and enlarged region of (b) valence band and (c) secondary electron cut-offs (SEC) in both the samples. (d) The enlarged view of valence band spectra showing a polaron peak just above VBM for undoped $V_2O_5$ (VONW) and the same being diminished for Mo doped $V_2O_5$ (MVONW)

Here, $Mo^{5+}$ is an isovalent impurity doping that could neither be an acceptor nor a donor. However, the presence of $Mo^{5+}$ affects the oxygen vacancy concentration as well as distribution. Besides, it subsequently affects the $V^{4+}$ ions concentration which is the primary charge hopping centre. It is observed that the VBM in the Mo doped sample are moved a little closer to $E_f$ than the undoped $V_2O_5$ as shown from Figure 3(b and d), the red lines intersecting the energy axis. For an n-type semiconductor like $V_2O_5$, this might signify the reduction in chemical potential i.e., effective $E_f$ being lower than that of undoped $V_2O_5$. This could have consequences in charge carrier hopping energies involved. This is further substantiated by the midgap states denoting a polaron peak just above the valence band edge seen in undoped $V_2O_5$, that diminishes after Mo doping as shown in Figure 3(d).

## C. Temperature dependent Electrical Transport studies.

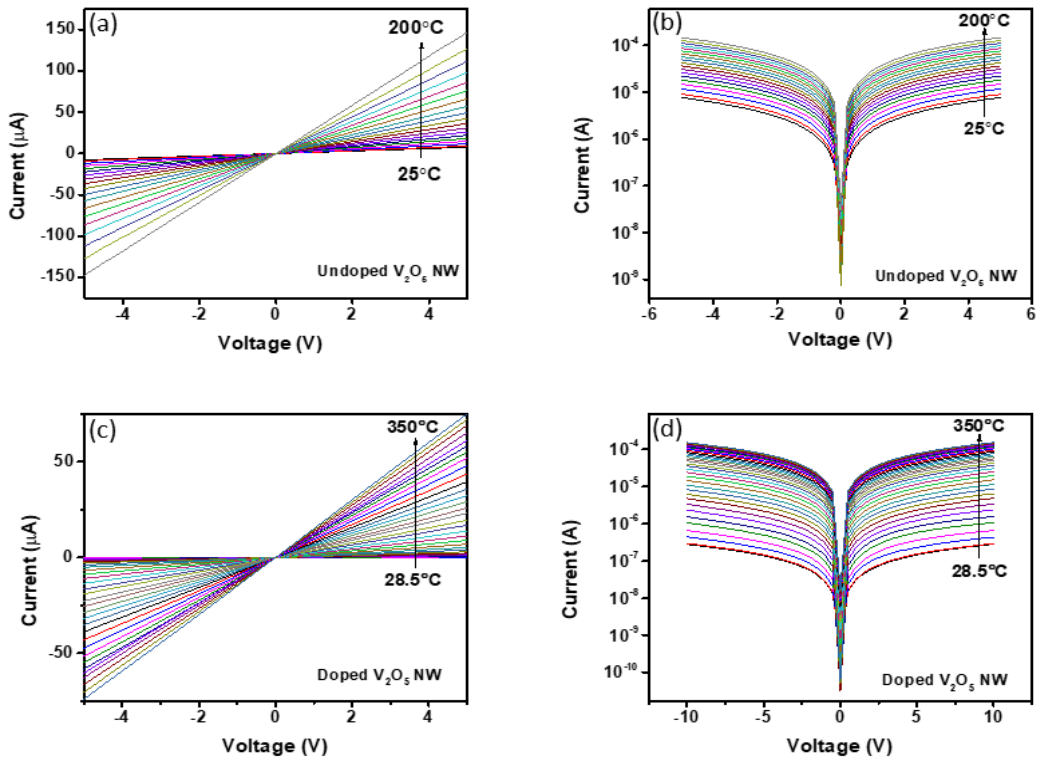

**Figure 4. The I-V characteristic in (a and c) linear and (b and d) log scale of undoped (upper row) and Mo doped $V_2O_5$ nanowires (lower row).**

The current–voltage (I-V) characteristics measured at different temperatures for VONW and MVONW are depicted in Figure 4. The I-V characteristics demonstrate a linear behaviour, confirming the ohmic contacts between the nanowire and the gold electrodes in either case. The work function difference is anticipated to be 100 meV between Au and VONW. The I-Vs of the MVONW also show linear behaviour.

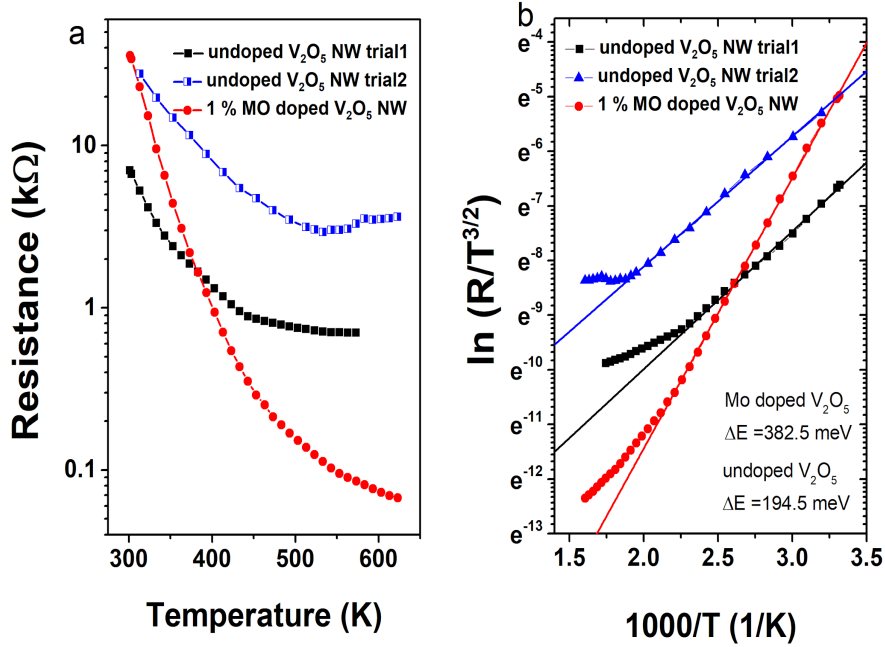

**Figure 5. (a) R-T graph of drop cast (undoped and doped) $V_2O_5$ nanowires at temperatures ranging from 25 °C – 350 °C (b) Corresponding analysis of R-T using polaron hopping model.**

In the R-T graph shown in Figure 5(a), the resistance decreased with increasing temperature indicating the typical semiconducting nature of the material. Two measurement trail data were recorded and the same are shown in Figure 5 as trail 1 and trials 2. It may be noted that although the nanowires R-T trend is the same irrespective of the room temperature R value. Both depict repeatable performance. In Figure 5(a), it may be observed that the resistance of the nanowires calculated from the slope of IVs, slightly increases after 260 °C (550 K). This has been attributed to the possible Metal-Insulator transition in literature [10,31]. It states that $V_2O_5$ undergoes a reversible fast transition from a semiconductor to metal-like near 257 °C. At higher temperatures, the band gap of $V_2O_5$ was found to decrease due to the spontaneous loss of surface oxygen and the formation of oxygen defects at the surface. However, the said behaviour is not confirmed by electron energy band changes and hence could only be surface chemical stoichiometry change. This results in enhanced non-stoichiometry of V ion resulting in metallic behaviour[31] due to $V^{3+}$, $V^{4+}$ in addition to $V^{5+}$. The change in slope of R-T data is evident of the transition. (See Fig S2 in supporting information section). In Mo doped sample, no such slope reversal is observed which again signifies that vanadium lower (+3, +4) oxidation states being less probable in doped nanowires.

The analysis using the Arrhenius equation gives a thermal activation energy of 194.5 meV and 382.5 meV for undoped and Mo-doped nanowires respectively when analyzed using Polaron hopping model[27]. It is depicted by eq (1).

$$R = R_o T^{3/2} \exp(E_{act}/k_B T) \qquad (1)$$

where, R and $R_o$ are the resistance values at variable temperature T and a reference temperature. Eact is the activation energy of polaron hopping and $k_B$ is Boltzmann constant. This represents energy for hopping between the energetically equivalent polaron sites which are formed due to lattice distortion near a vacancy site and having localization of electrons[9,32]. This extra electron can be localized on any of the neighbouring equivalent $V^{5+}$ ions making them $V^{4+}$. The presence of $Mo^{5+}$ substitution at one of the equivalent V ion sites, may reduce the polaron stabilization. As the oxygen vacancies and consequently the $V^{4+}$ ions together contribute to the polaron formation as mentioned in the XPS discussion, reduction in vacancies reduces the number of polarons and decreases their average relative abundance. This is reflected in high polaron hopping formation energy measured from slopes shown in Figure 5(b).

This results in higher separation among the polaronic sites which are contributed because of oxygen vacancy-induced lattice distortion and thereby localized electrons under coulomb attraction. In such a system, the charge transport happens through hopping among such polaronic sites and it gets thermally activated wherein the activation energy depends on the radius of polaron and inter-polaron separation. The lower the polaronic site density, the higher is the activation energy required for hopping. Thus, replacing some of the $V^{5+}$ ions with $Mo^{5+}$ ions reduces the equivalent sites for carrier hopping from $V^{4+}$ to $V^{5+}$ and affects the defect-related bands in the band picture. Besides, such reduction in mid-gap polaronic states can be directly visualized by UPS spectra near $E_F$ as shown in Figure 3. The reduction of polaron density would therefore be interpreted as a partial restoration of intrinsic conductivity and therefore lower carrier concentration.

### D. Gas sensor response studies.

The device fabricated from the nanowires samples and gold electrodes (as mentioned in the experimental section), is placed inside a gas sensor chamber which is equipped with a Mass flow meter based gas dilution system at the inlet and an exhaust. In a typical gas sensing experiment, the sample is maintained at a temperature of interest and timed pulses of different concentrations of given gas are injected. The response time is measured as the time taken by the sensor to reach a certain fraction of its saturation resistance value. A continuous flow of air is maintained at a constant flow rate (500 SCCM in this case) with or without the test gas. When the gas pulse is stopped, the sensor resistance drops/rises to its initial value and here the time taken by the sensor to reach its 80% of the saturation to baseline difference is taken as recovery time. Similarly, the concentration of the test gas is varied from 62-2163 ppm in each successive pulse.

Chemiresistive $V_2O_5$ sensors with ohmic properties show reversible change as a function of the concentration of ethanol vapours. Gas sensing performance is mainly evaluated on the basis of sensitivity, selectivity, stability, and the response and recovery tixmes of the sensing material[20].
For chemiresistive sensing, response (in percentage) is calculated by,

$$S(\%) = \frac{\Delta R}{R} = \frac{R_a - R_g}{R_a} \times 100 \qquad (2)$$

Where, $R_g$ – Resistance of the sensing material under target gas, $R_a$ - Resistance of the sensing material in air.

Many transition metal oxide sensors operate as gas sensors effectively at high temperatures. This can be attributed to the higher catalytic reaction rates as well as increased conductivity of metal oxides at high temperatures due to the formation of surface chemisorbed oxygen species. To measure the optimum sensor temperature, samples were first measured at increasing temperatures from 50 °C – 350 °C, keeping the gas concentration unchanged (1546 ppm). After finding the optimum temperature, measurements were taken with various concentrations of ethanol at that temperature. As temperature increases from 50 °C – 350 °C, the sensor response progressively increases with temperature and does not attain a maximum until 400 °C. Nevertheless, by 300 °C the response is significantly high. Therefore, it may be called the optimum temperature for maximum sensor response (as seen in Figure 6). below 300 °C the response drops sharply (shown in Figure 6) and also response times are fairly high at low temperatures. Above this temperature, the response may be higher. The Mo doped NW sample shows higher response for lower temperatures. On the other hand, the undoped nanowire have higher response at high temperature (>350 °C). This could be due to higher non-stoichiometry on undoped NW surface as seen in R-T data as MIT like transition around this temperature. While such transition could be seen in Mo doped NW and hence no such abrupt change in response may be expected as seen in undoped sample.

The response obtained is compared with that of literature reported values and the comparison is summarized in Table I.

Table I. Comparison of the response obtained in this study with that of the literature reports.

| $V_2O_5$ morphology | Dopant (if any) | Response Temperature | Response/Recovery time | Sensitivity value | Target gas concentration | Reference |
|---|---|---|---|---|---|---|
| Nanorods | - | 25 °C | 20s/27s | 3.5% | 500 ppm | 33 |
| Nanobelts | - | 200 °C | 30s/50s | 68% | 1000 ppm | 34 |
| Thin films | - | 250 °C | 5s/8s | 1.25% | 500 ppm | 35 |
| Nanoflowers | - | 300 °C | 25s/14s | 3.3% | 100 ppm | 36 |
| Nanostructures | Ag | 250 °C | No data | 62% | 100 ppm | 37 |
| Nanostructures | In | 250 °C | No data | 56% | 100 ppm | 37 |
| $V_2O_5$ nanowires | - | 300 °C | $T_{res}$= 1 min to 80% sat $T_{rec}$= 23 min to 20% sat | 13% | 1546 ppm | This work |
| $V_2O_5$ | Mo | 300 °C | $T_{res}$= 1.5 min to 80% sat $T_{rec}$= 16.9 min to 20% sat | 23.5% | 1546 ppm | This work |

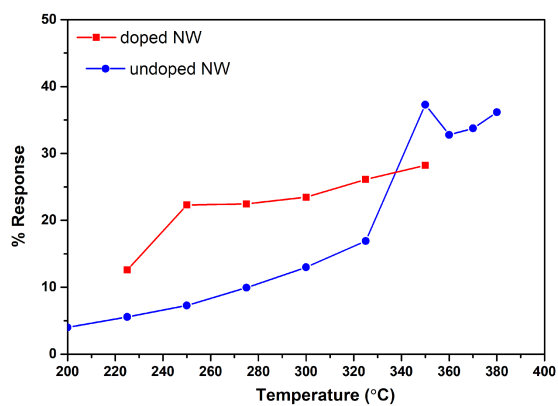

**Figure 6. The comparison of response of undoped and Mo doped $V_2O_5$ nanowires for the same concentration of ethanol as a function of temperature for 1546 ppm of ethanol vapors.**

The response at a given temperature for different concentrations of ethanol vapours is studied at 300 °C and the same is shown in Figure 7. There is a sudden drop in the resistance with exposure to gas and the sensor shows another slow fall to saturation. Similarly, upon withdrawing the test gas (purging only air), the sensor signal recovers initially fast and followed by a slow recovery. In case of the doped sample not only is the sensitivity found to be higher for a given concentration, but the response and recovery times were also found to be improved substantially as seen in Figure 7 (a) and (b).

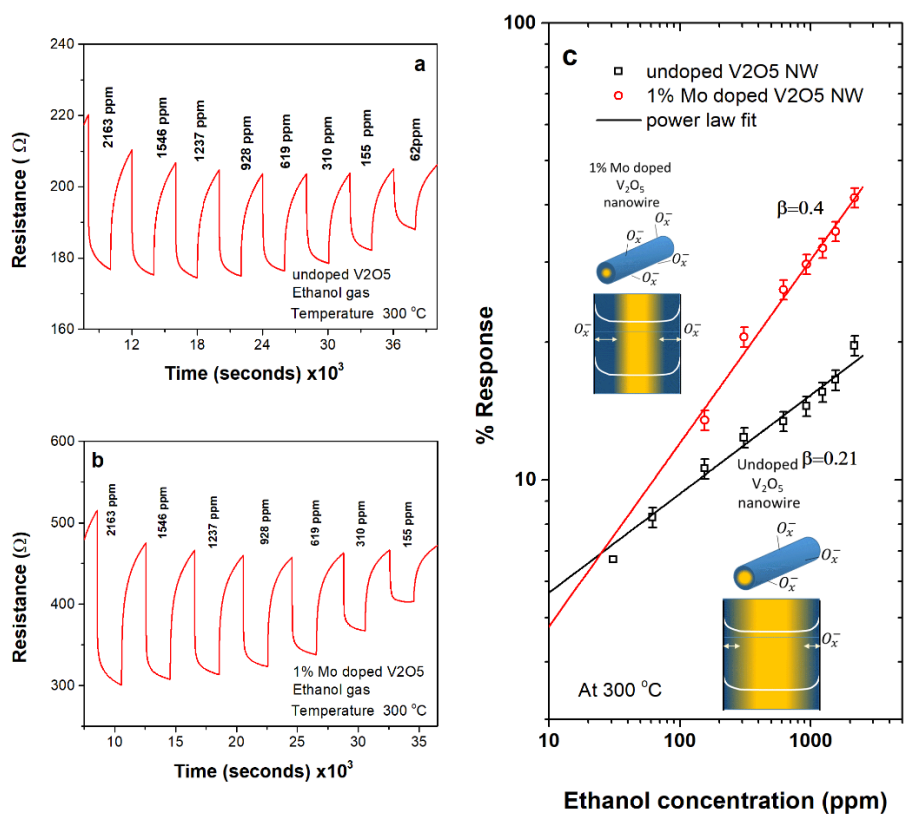

**Figure 7. Gas sensing response of drop cast (a) undoped and (b) 1% Mo doped V$_2$O$_5$ nanowires with ethanol as target gas at different concentrations (c) Power law fit for the sensing response data. The schematic diagrams of the radial band bending for the nanowires are shown in insets for undoped and 1% Mo-doped V$_2$O$_5$ nanowires.**

The response vs concentration is found to obey the well-known power law behaviour as shown in Figure 7(c). As the temperature is increased, the number of adsorbed oxygen on the surface increases resulting in an increase in sensitivity[23]. Figure 7(c) depicts that the nanowires respond to ethanol concentration levels as low as 50 – 430 ppm at 300 °C. At this temperature, the response of Mo doped nanowires was substantially larger than the undoped nanowires for the concentration range studied. Extending the concentration of vapours to a lower concentration is anticipated that a crossover of sensitivity may occur below about 20 ppm.

The response time and recovery times of the sensor are strong function of concentration in addition to the temperature and material morphology. In general, the response times and recovery times decrease as temperature is increased, due to thermal activation of the catalytic reaction. Metal oxides are good catalyst where the oxide sensor surface reduces or oxidises the gas efficiently at optimally high temperatures[38].

Here the gas sensor response as well as recovery times were compared for VONW and MVONW for a fixed gas concentration (1546 ppm ethanol). Typically, response (and recovery) times are estimated as a time taken by sensor signal to rise to (or fall to) certain percentage of the signal saturation value. The times taken to reach 80% of signal saturation and fall to 20 % of the same are listed in Table 1. The Mo doped sensor shows a faster timescale. Nevertheless, most of the sensor signal rises in first few seconds and is followed by a slow rise thereafter. Similarly, for the recovery. Thus we envisage there are two time constants involves in the response as well as recovery and those are having an exponential contribution as expected for solid state oxide gas sensors. The two time constants $T_1$ and $T_2$ may arise from surface as well as bulk (diffusion) reactions and the same are evaluated using eq (3) and (4)[39] for response,
$$G(t) = G(0)\left[1 - \exp\left(\frac{t}{T_1}\right)\right] + G'(0)\left[1 - \exp\left(\frac{t}{T_2}\right)\right] \quad (3)$$

For recovery,
$$G(t) = G(0)\exp\left(\frac{-t}{T_1}\right) + G'(0)\exp\left(\frac{-t}{T_2}\right) \quad (4)$$

Where, G(t), G(0) and G'(0) are the dynamic and saturation value of conductances for response and recovery.

The two values obtained for $T_1$ in either case are about 10 sec while the $T_2$ are ~500 sec for response while the same are 100 and 1000 for recovery, as shown in Fig 8 (a and b).

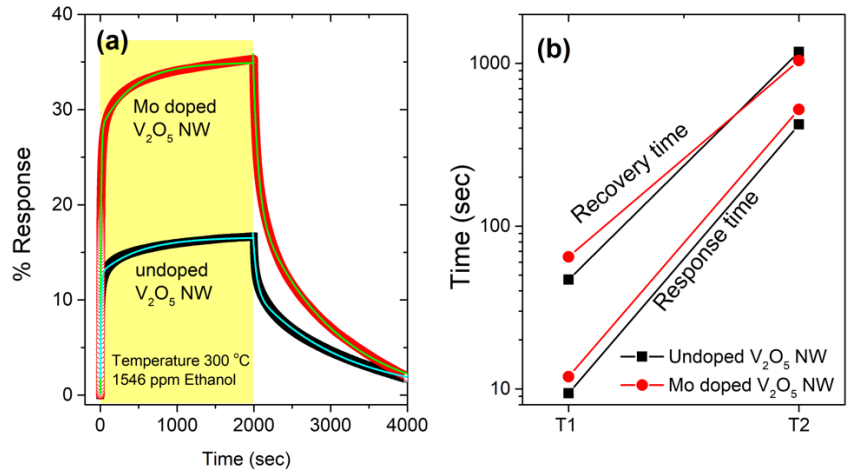

**Fig 8. (a) The response and recovery times for 1546 ppm ethanol vapors at 300°C with fit using eq (3 and 4). (b) The time constants $T_1$ and $T_2$ calculated.**

The sensitivity of $V_2O_5$ (like other SMOs) is attributed to the formation of adsorbed oxygen on its surface. The sensing body is a porous assembly of tiny grains of metal oxide. On exposure to air, oxygen gets adsorbed on the grains as anionic species. Here it is shown by the $(-o_x^-)$ which indicates the surface adsorbed oxygen species. The x value of 1 stands for atomic species and 2 for molecular oxygen species. However, they carry a negative charge due to electron capture from oxide. Typically atomic O¯ exists at temperatures greater than 200 °C and induces an electron-depleted layer to increase the surface potential and work function. The charge carrier densities are extremely low in these depleted regions and the same is shown as band bending in the band diagram. In cases where the grains are connected to their neighbours through the grain boundaries, a potential barrier for the migration of electrons, often called a double-Schottky barrier, is formed across each grain boundary. This barrier plays a dominant role in determining the resistance of the sensing body. Adsorbed oxygen will oxidize a reducing gas leading to mitigation in the potential barrier and then also in resistance of $V_2O_5$. On the other hand, in contact with an oxidizing gas in air, the interaction between adsorbed gas and oxygen will withdraw electrons from the semiconductor core, thus increasing resistance.

The response of the doped $V_2O_5$ is better than the undoped nanowire sensor which may be attributed to the reduction of $V^{4+}$ ions in the doped sample resulting in lesser hopping sites and lower carrier density. The lower carrier density results in higher surface carrier depletion width, (W) as well as band bending in the air as shown in Figure 9. As the reducing gas is adsorbed, the carriers are donated to the oxide, resulting in the lowering of depletion width as well as band bending. The depletion width (W) is given by,

$$W = \sqrt{\frac{\varepsilon \varepsilon_r V_s}{q.n}} \tag{5}$$

where, $\varepsilon_r$ is the dielectric constant, $V_s$ is the surface band bending potential, and n is the carrier concentration and q is the electronic charge. Therefore, in a Mo doped sample if n reduces, the depletion width, W would increase.

This may result in a larger change in resistance in the case of doped nanowires as compared to undoped ones. Thus, this enhancement can be attributed to the reduction of oxygen vacancies and $V^{5+}$ sites, which reduces the hopping sites of those major conduction mechanisms in vanadium pentoxide. The replacement of $V^{5+}$ with $Mo^{5+}$ may reduce overall electrical conductivity in two ways; one is the isovalent doping resulting in no new charge carrier addition and the other is the reduction in hopping sites. Besides, the Mo ions act as active sites for binding oxygen from the air and also the test gas molecules, hence their presence may induce catalytic contribution to sensor response in addition to electronic contribution.

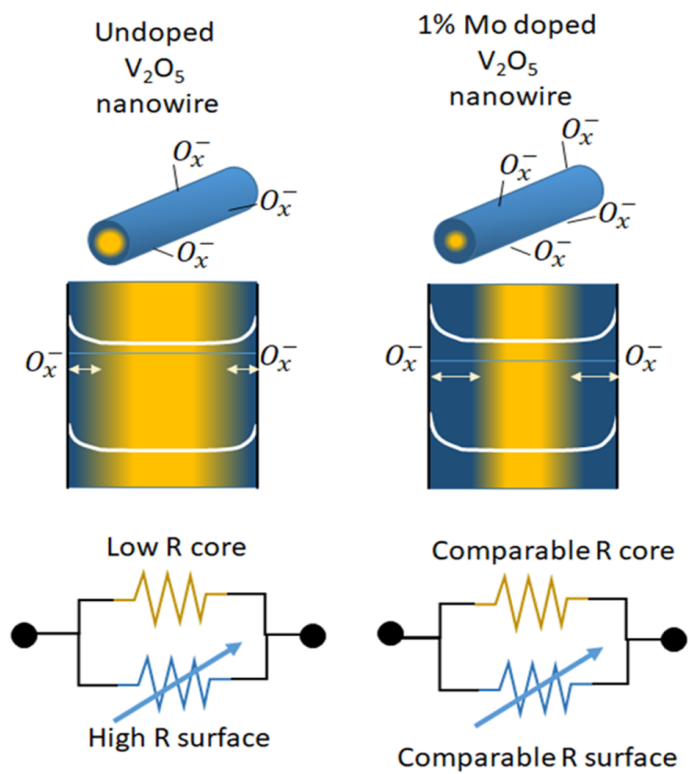

**Figure 9.** The schematic diagram showing the radial surface depletion profile for undoped and Mo doped $V_2O_5$ nanowires, their band bending as well as equivalent circuits for parallel core and surface resistors. The $-o_x^-$ indicates the surface adsorbed oxygen species where x could be 1 for atomic species and 2 for molecular oxygen species. However, they carry a negative charge due to electron capture from oxide. The yellow region represents flat band condition (no electric field) whereas the blue region represents surface band bending change in potential.

IV.     Conclusion

To summarise, we have synthesised $V_2O_5$ nanowires with and without Mo dopants and having a long aspect ratio as the sensing material. The material characterization using Scanning Electron Microscopy (SEM) and X-ray diffraction (XRD), and electrical characterization by plotting the resistance change with respect to temperature show that the nanowires obtained are sufficiently long and are conducting (a few kΩ resistance at room temperature). The XRD data shows that Mo is substituted at the V site and no impurity phases are observed. The XPS studies confirm the 5+ oxidation state of Mo ions and reveal a reduction in $V^{4+}$ ions as well as oxygen vacancies as a result of Mo substitution. The introduction of Mo sensing analysis concludes that both doped, as well as undoped $V_2O_5$ NWs, are n-type semiconductors having small polaron hopping types of conduction and the hopping energy is higher in Mo-doped NW. The undoped VONW shows insulator to metal-like transition beyond 550 K. This is also reflected in sensor response as a sudden jump in response value. Whereas the same is not found to be seen in MVONW which again signifies relatively lesser reduced forms of V ions that bring about such transition. MVONW are found to show better response to ethanol vapours at lower temperatures which is ascribed to suppressed polaronic conductivity, resulting in lowered carrier concentration and subsequently higher depletion width. As a result, the sensing measurements performed using ethanol as the target gas show that Mo doping has enhanced the sensing response compared to the undoped sample. Further, studies are ongoing on the temperature dependence of sensitivity and carrier type reversal at high temperatures in doped as well as undoped $V_2O_5$ nanowires.

**Supplementary Material**

The Supplementary Material includes the enlarged views of various X-ray diffraction lines for doped and undoped $V_2O_5$ nanowires and the first derivative of R vs T depicting presence of MIT in undoped $V_2O_5$.

**Acknowledgment**
The authors are thankful to DST Nano Mission (DST/NM/NT/2018/124) and SERB core research grant (CRG/2022/006973) Govt. of India for the funding support received. The Central Instrumentation Facility of IISER Thiruvananthapuram is also acknowledged for the XPS, XRD and SEM facilities.

**Data availability statement**
The data is available with the corresponding author upon reasonable request.

**Conflict of Interest**
The authors declare no conflict of interest.